\documentclass[journal]{IEEEtran}
\usepackage{amssymb}
\usepackage{amsmath}
\usepackage{graphicx, subfigure}
\usepackage{url, cite}
\usepackage{verbatim}
\usepackage{booktabs}
\usepackage{multirow}
\usepackage{bigstrut}
\usepackage{booktabs}
\usepackage{amsmath,cite,graphicx,times,subfigure,url,verbatim}
\usepackage{algorithmic}
\usepackage{algorithm}
\usepackage{epstopdf}

\newtheorem{subsec:coding}{subsec:coding}

\usepackage{color}

\begin{document}


\title{Wearable Affective Robot}

\author{
Min~Chen,~Jun~Zhou,~Guangming~Tao,~Jun~Yang,~Long~Hu

\thanks{M. Chen is with the Wuhan National Laboratory for Optoelectronics and the School of Computer Science and Technology, Huazhong
University of Science and Technology, Wuhan 430074, China, and also with Wuhan AIWAC Robotics Co., Ltd, China. 
Email: minchen@ieee.org}

\thanks{J. Zhou is with the Wuhan National Laboratory for Optoelectronics and the College of Optoelectronic Science and Engineering, Huazhong University of Science and Technology, Wuhan 430074, China.
E-mail: junzhou@hust.edu.cn}

\thanks{G. Tao is with the Wuhan National Laboratory for Optoelectronics and the School of Optical and Electronic Information, Huazhong University of Science and Technology, Wuhan 430074, China.
Email: tao@hust.edu.cn}

\thanks{J. Yang is with the School of Computer Science and Technology, Huazhong University of Science and Technology, China and Wuhan AIWAC Robotics Co., Ltd, China. 
Email: junyang\_cs@hust.edu.cn}

\thanks{L. Hu is with Wuhan Emotion Smart Sensing Technology Co., Ltd, China. 
Email: hulong@hust.edu.cn}

\thanks{Jun Yang is the corresponding author.}
}

\markboth{Under Proof: IEEE Access, VOL. XX, NO. YY, MONTH 20XX}{}

\maketitle

\begin{abstract}
With the development of the artificial intelligence (AI), the AI applications have influenced and changed people's daily life greatly. Here, a wearable affective robot that integrates the affective robot, social robot, brain wearable, and wearable 2.0 is proposed for the first time. The proposed wearable affective robot is intended for a wide population, and we believe that it can improve the human health on the spirit level, meeting the fashion requirements at the same time. In this paper, the architecture and design of an innovative wearable affective robot, which is dubbed as Fitbot, are introduced in terms of hardware and algorithm's perspectives. In addition, the important functional component of the robot-brain wearable device is introduced from the aspect of the hardware design, EEG data acquisition and analysis, user behavior perception, and algorithm deployment, etc. Then, the EEG based cognition of user's behavior is realized. Through the continuous acquisition of the in-depth, in-breadth data, the Fitbot we present can gradually enrich user's life modeling and enable the wearable robot to recognize user's intention and further understand the behavioral motivation behind the user's emotion. The learning algorithm for the life modeling embedded in Fitbot can achieve better user's experience of affective social interaction. Finally, the application service scenarios and some challenging issues of a wearable affective robot are discussed.
\end{abstract}


\section{INTRODUCTION}\label{sec:introduction}

Artificial Intelligence (AI) is defined as the ``machine intelligence imitating human behaviors and cognitive abilities''~\cite{Cisco2014Global}. In recent years, the AI has received continuous concerns from both academia and industry, and various countries have invested in the AI-related research and development. The data obtained from China Industrial Economic Information Network (CINIC) show that the global investment in artificial intelligence has been grown from 589 million US dollars in 2012 to more than 5 billion US dollars in 2016. It is estimated that by 2025, the market capitalization of AI applications will reach 127 billion dollars~\cite{Chen2016Mobility}. With the continuous development of AI technology, the AI and medical health field have been integrated, which has formed an important interdiscipline (AI-based medical treatment) closely related to the national economy and people's livelihood~\cite{chen5G-Cognitive}. According to the forecasts, by 2025, the AI-based medical treatment industry will occupy one fifth of the market scale. Although the AI and Medical treatment in China has started a little later than in some other countries, it is expected that it will reach 20 billion Chinese Yuan of the market size in China by 2018.

The AI-based medical treatment interdiscipline plays an important role in the development of the artificial intelligent-based human health diagnosis and treatment~\cite{Zheng2017}~\cite{chenEdgeCognitive}. In 2011, it was found by the research personnel from the Langone Health of New York University that, the analysis and matching of the pulmonary nodule images (chest CT images) based on the AI are 62\% to 97\% faster than the manually-annotated ones conducted by the radiologists, which could save up to 3 billion dollars every year. Another research on 379 patients in the Plastic Surgery field showed that in comparison with the independent operations performed by surgeons, the AI robot-aided technology created by Mazor Robotics reduced the surgical complications fivefold, which could reduce 21\% of the patient's post-operation hospital stay, which further could bring less pain and faster recovery to patients, providing more effective guarantee to patient's healthy life and saving up to 40 billion dollars each year~\cite{SPHA}.

In comparison with the physiology, the psychology is a more important factor guaranteeing the human health and life happiness. When our environment is short of the experience of stability or belongingness, we may reproduce a pleasant emotion through television, movie, music, book, video game, or any other thing which can provide an immersive social world~\cite{wearableGhasemzadeh}~\cite{CognitiveZhang}~\cite{GeJointOptimization2018}. The basic emotions of humans are well-founded, even in the form of the virtual artificial intelligence, such as in a virtual assistant, a traditional service robot, etc. Among them, the virtual assistant uses the Natural Language Processing (NLP) to match the user text or voice input with its executable commands, and continuously studies them by the AI technologies including the machine learning. On the other hand, the traditional service robots (without a virtual assistant), such as sweeping robots and industrial robots, only provide the mechanical services~\cite{leiDelayoptimal2016}~\cite{YangDense2018}~\cite{zhengSoft-defined2016}.

However, humans have a stronger emotional response to the tangible AI~\cite{emotionChen}. In other words, the more human-like a robot is, the stronger emotional response we will have to it. The intelligent robots are mainly divided into social robots, affective robots, and wearable robots.
\begin{itemize}
\item Combining a traditional service robot with a virtual assistant, a social robot, which has the ability to imitate one or multiple cognitive competencies, such as natural language interaction, is created. The traditional social robot is a type of the social robots. Through a mobile application, it can interact with humans and other robots. A social robot owns a virtual assistant and has the mechanical ability, such as the ability to move its body parts~\cite{Zhou2018}~\cite{iotSmartHealthcare}~\cite{energySDNwsn}. Current social robots focus on four major applications: health care and treatment, education, public and work environment, and home environment~\cite{M2MZhang}~\cite{jiangHome}~\cite{smartHome2.0}. Although the social robots have the language-based communication ability and can imitate human behaviors, they do not have strong intellectuality, such as cognition of human emotion~\cite{autoMethodsGravina}.
\item On the other hand, an affective robot has the function of emotion recognition from the third-person perspective, but it cannot accompany users conveniently because its human-computer interactive mode is limited by the surrounding environment.
\item In recent years, there is the concept of a wearable robot in Cybathlon organized by ETH. However, it actually refers to the exoskeleton of a robot. Currently, its application scope is limited to the disabled people. We believe the application range of a wearable robot should be enlarged to various population.
\end{itemize}

\begin{figure}
\centering
\includegraphics[width=3.5in]{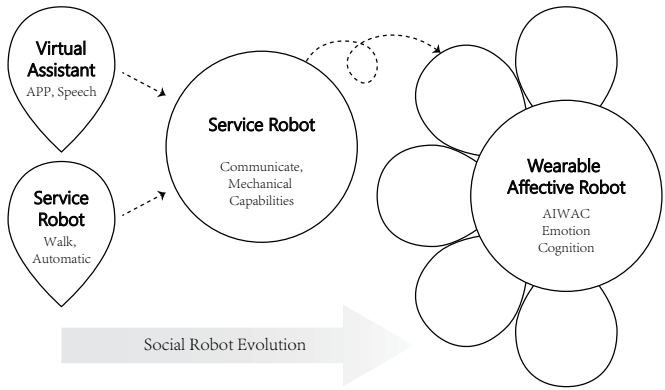}
\caption{The Evolution of Wearable Affective Robot}
\label{fig01}
\end{figure}

In general, the research on social affective robots is still in the preliminary stage, and its development difficulties mainly refer to the following aspects.
\begin{itemize}
\item \textbf{Cost:} A social affective robot currently costs about a few hundred to a few thousand dollars, which is difficult to afford.
\item \textbf{Intelligence:} Currently the research on the machine intelligence is still in a non-mature stage, and the cognitive competence of humans has not been achieved yet, so human emotion cannot be recognized accurately.
\item \textbf{Human-computer interaction interface:} The interface is not friendly enough, especially that of the hard robot shell which limits the barrier-free communication between humans and machines.
\item \textbf{Movability:} Users moves in a large range in the physical world, but current social affective robots do not have such a large-range movability.
\item \textbf{Comfort:} Current social robot or robot exoskeleton has a great disturbance to users, and cannot provide the close-fitting and comfortable experience.
\item \textbf{Service object:} At present, the focus is on patients or special groups without the possibility to cover the overall population. These shortcomings limit the use of social emotional robots, so the new-generation robots should have more comprehensive and powerful functions, and their application scope should cover a wider population~\cite{ZhengSMDPbased2015}~\cite{SPINE2}~\cite{mobilityFortino}.
\end{itemize}

Here, we present the problems that must be solved to overcome the challenges arising from the above-mentioned shortcomings.
\begin{itemize}
\item The third-person perspective and first-person perspective should be switched over flexibly.
\item It should be integrated with the user comfortably without affecting the social contact and daily traveling of a user.
\item The personalization must be realized, the in-depth modeling of a user should be accomplished, and current decision-making should be influenced through a long-term cognition process on a user, so as to improve the robot intelligence continuously.
\end{itemize}

Therefore, in this work, an intelligent affective interaction robot, an intelligent haptic interaction device, and an intelligent brain wearable device are combined to realize a new Fitbot in the form of smart clothes. This is a brand new morphology of a social affective robot, which simulates the thinking process of humans based on a cognitive computing model. Moreover, it involves the utilization of data mining, pattern recognition, natural language processing, and other machine learning techniques to simulate the working mode of a human brain, which allows robots to interact with users in a more complex way. The proposed Fitbot can improve the accuracy of user's emotion recognition by the acquisition of breadth, long-term, and in-depth data of a single-user, as well as by using the unlabeled learning, emotion recognition, and interactive algorithm. A Fitbot can significantly influence the actual life of human beings from the aspect of spiritual life, emotional care, medical rehabilitation, intelligent company, etc.

The evolution process of Fitbot is presented in Fig.~\ref{fig01}, a virtual assistant is a pure software application, while a service robot can move autonomously, but has no language skill. Thus, a service robot can be combined with a virtual assistant forming a social robot with the ability to communicate using the language. However, a social robot is not necessarily intelligent and can only simulate human language. To further develop the intelligence of a social robot, the human feelings must be recognized, which leads to a social affective robot. However, a social affective robot does not have high comfort and personalization, so a Fitbot is proposed in this paper. By introducing a strong AI, a Fitbot is endowed with the ability to cognize the emotion, having the portable and fashionable elements, so as to serve people in a wider range. Moreover, it can improve the people ``health'' on the spirit level. Therefore, a Fitbot is a social affective robot whose ability to cognize emotions can be improved on its own. Table \ref{tab01} gives the performance comparison between the Fitbot, smart clothes, social robot (taking a chatting robot as an example), wearable robot, and affective robot.

\begin{figure}
\centering
\includegraphics[width=3.5in]{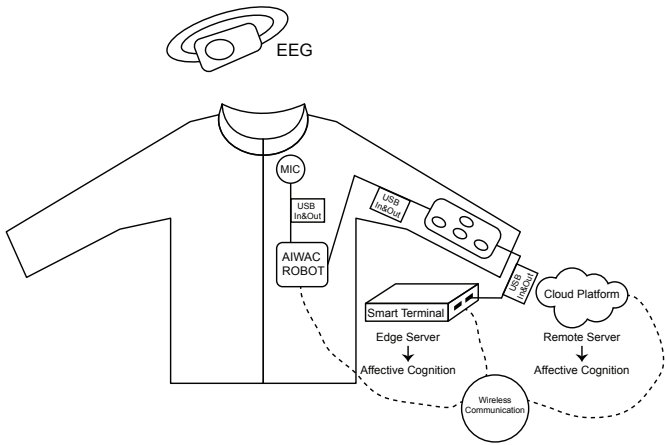}
\caption{The architecture and constituent parts of a Fitbot}
\label{architecture}
\end{figure}

\begin{table*}
\renewcommand{\arraystretch}{1.2}
\caption{Comparison of different robots }
\label{tab01} \centering
\begin{center}
\newcommand{\tabincell}[2]{\begin{tabular}{@{}#1@{}}#2\end{tabular}}

\begin{tabular}{|c|c|c|c|c|c|c|c|c|c|c|}
\hline
\textbf{\tabincell{c}{Product\\ Name}} & \textbf{Category} & \textbf{Sociality} & \textbf{\tabincell{c}{Comfort\\ Index}}  & \textbf{Usability}  & \textbf{Washable}  & \textbf{Accuracy} & \textbf{Sustainability} & \textbf{\tabincell{c}{Physiological\\ Index}} & \textbf{Real-time} & \textbf{Consumers}\\
\hline
 \tabincell{c}{Smart \\ clothing} & Wearable 2.0 & No & High & Very easy & Yes & Middle & Yes & Complex & Yes & Everyone\\
\hline
Bots & \tabincell{c}{Social \\Affective \\Robot 1.0 }& Simple & Middle & Easy & No & Middle & Yes & No & Yes & Everyone\\
\hline
\tabincell{c}{Wearable \\Robot }& \tabincell{c}{Social \\Affective \\Robot 1.0 }& No & Low & Hard & No & Low & Yes & Simple & No & Deformity\\
\hline
\tabincell{c}{Affective \\Robot} & \tabincell{c}{Social \\Affective \\Robot 1.0} & Complex & Middle & Easy & No & Middle & Yes & No & Yes & Everyone\\
\hline
 \tabincell{c}{Wearable\\ Affective\\ Robot} & \tabincell{c}{Social \\Affective \\Robot 2.0} & \tabincell{c}{Very\\ Complex} & High & Very easy & Yes & High & Yes & \tabincell{c}{Very\\ Complex} & Yes & Everyone\\
\hline
\end{tabular}
\end{center}
\end{table*}

In general, the main contributions of this paper are as follows:

\begin{itemize}
\item A wearable robot having the emotion-cognition ability named the Fitbot is presented. The composition and structure of the presented robot are introduced in detail from the aspect of hardware design and algorithm design.
\item The important constituent part of a Fitbot, the brain wearable device, is introduced in detail. The innovation research in the brain wearable field is introduced from the aspects of hardware design, EGG data acquisition and analysis, user behavior perception, algorithm deployment, etc.
\item Using the cognitive computing, and through the continuous data acquisition in depth and breadth, the life modeling of a user is achieved, which truly realizes a personalized intelligent Fitbot.
\item Finally, the application services and the market prospect of a Fitbot are summarized, and the future challenges are discussed.
\end{itemize}

The remaining of this paper is organized as follows. In Section~\ref{sec:architecture}, a Fitbot architecture is introduced. In Section~\ref{sec:hardware}, the hardware design of a Fitbot is presented. In Section~\ref{sec.algorithmn}, the algorithm design of a Fitbot is explained. In Section~\ref{sec.life-modeling}, the process of life modeling is introduced in detail. In Section~\ref{sec.application}, the applications and services of a Fitbot are discussed. Lastly, in Section~\ref{sec.conclusion}, the challenges that Fitbot faces are listed, and the paper is concluded.

\section{FITBOT ARCHITECTURE AND DESIGN ISSUES} \label{sec:architecture}

\subsection{Fitbot Architecture}

As previously mentioned, the proposed Fitbot integrates the intelligent affective interaction robot, intelligent haptic interaction device, and intelligent brain wearable device, and represents a new-type wearable robot with the emotion-cognition ability in the way of smart clothes. The architecture and constituent parts of a Fitbot are shown in Fig.~\ref{architecture}.

On the basis of the fashionable clothes commonly worn by people, we have designed an isolation layer for storing the smart terminal (a smartphone) according to the physiological structure of a human body and using the users' habits, so as to facilitate its physical connection and interaction with other modules such as AIWAC smart box, AIWAC smart tactile device, brain wearable device, AIWAC robot, and cloud platform. The smart terminal can be considered as the edge server of the system~\cite{Chen2016Efficient}, which has the function of affective cognitive. Different application scenarios can be realized using a smartphone. In addition, we integrate the design concept of the Wearable 2.0, i.e., the AIWAC smart tactile device. We physically connect the smart tactile device with the AIWAC device and smartphone (via USB I/O mode), which can perform simple control operations and tactile perception~\cite{Tao2012}~\cite{Tao2016}, so as to realize the tactile human-computer interaction process. Inside the smart clothes, we deploy a Fitbot using the AIWAC smart box as hardware, which can provide the visual and voice interactions with users through the wireless communication of a smartphone and an external MIC device. Due to the physical connection with the smart tactile device via USB I/O mode, a Fitbot possesses rich tactile perception and interactive functions simultaneously. In addition, the AIWAC smart box integrates the local emotion recognition algorithms and has the off-line affective recognition and interaction functions. When there are relatively high requirements for the affective recognition, data access, and service quality, an AIWAC smart box may also skip the smart terminal(edge server)~\cite{Wang2014Mobility}~\cite{wangEdge}, and make direct wireless communication with the cloud platform (equivalent to the system's remote server)~\cite{chenNarrow}~\cite{minCognitive}~\cite{chenSOVCAN}, so as to realize more accurate and diversified affective cognitive services.
Taking emotion as a communication element to transmit between Fitbot and cloud platform, we offer a better solution to resolve
the multidimensional modal emotional communication problems for Fitbot, so as to provide users with personalized emotional service.

To better percept and analyze users' emotions in a multi-modal way~\cite{minDeep}, we integrate the intelligent brain wearable devices in the cap of the smart clothes, which can collect the users' EEG (Electroencephalograph) data. The collected EEG data is transmitted to the background through the wireless communication, so as to realize the perception of users' subtle expression changes. On the one hand, the detection of these subtle behaviors can ensure perception and analysis of the user's behavior patterns; on the other hand, some high-efficiency human-computer interaction patterns may be perceived by virtue of these accurate behaviors to achieve good emotional interaction experience.

\subsection{Design Issues}

Although a Fitbot can provide customized services for users, which greatly improves the user experience, it also brings many problems such as private information leakage, duration, network reliability, etc.

\begin{itemize}
\item \textbf{Problem of privacy invasion:} Currently, this is the most concerned problem because it harms the users greatly. Since a huge amount of extensive and in-depth data of users is collected, the users' privacy exposure will be very serious if the data leakage occurs. On the other hand, the organizations such as cybercriminals may continuously monitor the user environment and collect sensitive data of a user. If the user data are acquired by these criminals, it can also cause great harm to the user~\cite{SmartHomeCicirelli}~\cite{zhaoILLIA}~\cite{Chen2015On}.
\item \textbf{Problem of duration:} Because of the movability, a Fitbot cannot be connected to the power supply continuously. Thus, when the Fitbot battery is used up, the user with a high dependence on a robot will be affected, so the product experience will be seriously impacted. Therefore, the optimization of the energy consumption is a key problem that needs to be solved urgently~\cite{Xue2018}~\cite{edge-CoCaCo}~\cite{demandMaharjan}.
\item \textbf{Problem of network reliability:} This problem mainly refers to the reliability of cloud services. In the case of network interruption, the analysis algorithms and other intensive computing dependent on a social robot will not work. A possible solution to this problem is to design a flexible data processing algorithm, which can perform computing locally without a need for network connection~\cite{Zhang2016Energyaware}~\cite{chenTaskOffloading}~\cite{Liu2016Delay}. Another feasible solution is to develop a machine learning chip and integrate that chip into small devices to realize the data processing on a device (such as the A11 bionic chip used by Apple and the Movidius neuron computing bar used by Intel).
\end{itemize}

\section{DESIGN OF WEARABLE AFFECTIVE ROBOT HARDWARE}  \label{sec:hardware}

A Fitbot is mainly formed by three types of intelligent hardware: AIWAC smart box, brain wearable device, and AIWAC smart tactile device with the combination of the Wearable 2.0 technology. A Fitbot uses the AIWAC smart box as a hardware core for service processing to execute the data persistence functions such as service data receiving, storage, transmission, etc. A Fitbot also deploys the corresponding affective recognition algorithm relying on the service data. A Fitbot integrates the brain wearable device to achieve the perception of the users' EEG data, which are transmitted to the cloud platform through the AIWAC smart box for processing and analysis, and the cloud platform realizes the perception of users' affective state through the analysis of the EGG data. The AIWAC smart tactile device is flexibly integrated into a Fitbot in the PnP mode, and it allows a Fitbot to have the comfortable haptic interaction ability by combining it with the Wearable 2.0 technology, so as to enrich robot's human-computer interaction interface.

\subsection{AIWAC Smart Box Hardware Design}
\begin{figure}
\centering
\includegraphics[width=3.5in]{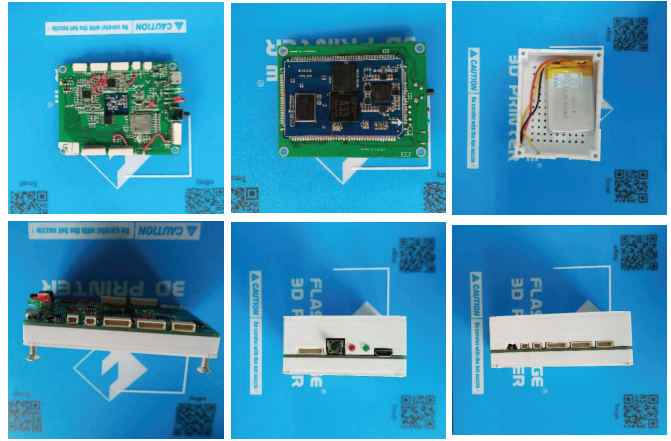}
\caption{The AIWAC smart box hardware design}
\label{fig02}
\end{figure}

The AIWAC smart box can make visual and voice interaction, providing the rich tactile perception and interaction functions. In addition, the AIWAC smart box integrates the local emotion recognition algorithms, and has the off-line affective recognition and interaction functions. Relying on the affective recognition and interaction system, by integrating the AIWAC smart box, a Fitbot can have nine types of personality characteristic including brave, steady, sincere, kind-hearted, self-confident, tenacity, forward-looking, and optimistic, and can recognize 21 different human emotions. The AIWAC smart box is the important hardware part for realizing the user emotion pacifying and health regulation, the profile of AIWAC smart box is shown in Fig. \ref{fig02}.

\subsection{Brain Wearable Hardware Design}

\begin{figure}
\centering
\includegraphics[width=3in]{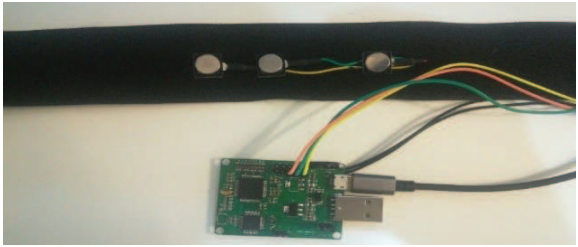}
\caption{The picture of the EEG device}
\label{fig0303}
\end{figure}

The cerebral cortex of a human can produce weak electrical signals which contain a large amount of information. However, it is a challenge to collect the weak electrical signals from the cerebral cortex while removing the numerous noise signals in the natural environment. After the profound investigation, as shown in Fig.~\ref{fig0303}, we design a 1-Channel EEG device for data acquisition, which consists of three electrodes connected in series, the IN1P EEG signal collecting electrode, refer reference electrical signal electrode, and BIAS1 paranoid drive electrode.

\subsection{AIWAC Smart Tactile Device}

To realize a better interaction of a Fitbot with the users, we have designed a haptic extension interaction using the AIWAC smart tactile device, which is not limited to the traditional voice interaction mode. We design the PnP mode to integrate the AIWAC smart tactile device into a Fitbot or other intelligent devices, so as to carry out the basic interactive control operations by using the AIWAC smart tactile device, and further achieve the convenient human-computer interaction using a Fitbot.



\subsection{Wearable Affective Robot prototype }

\begin{figure*}
\centering
\includegraphics[width=5in]{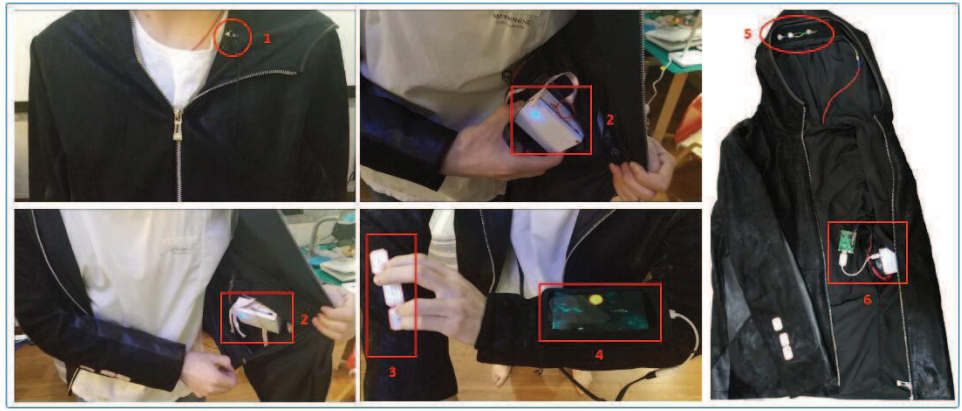}
\caption{The prototype of a wearable affective robot}
\label{fig0501}
\end{figure*}

A Fitbot integrates the AIWAC smart box, AIWAC smart tactile device, and AIWAC brain wearable device, and its prototype is shown in Fig. ~\ref{fig0501}, where position 1 denotes the voice interaction module of the AIWAC smart box, position 2 denotes the hardware core of the AIWAC smart box, and position 3 denotes the AIWAC smart tactile device integrated into the Fitbot. As shown in position 5, the brain wearable device is integrated into the cap position of a Fitbot, and its mainboard is connected with the rest of a Fitbot via USB communication, which is shown in position 6. A Fitbot has the rich means of perception and interaction, such as voice perception, tactile perception, EEG perception, and etc. On this basis, it achieves the multi-modal perception and analysis of users' emotions and achieves good affective interaction experience.

\begin{figure*}
  \centering
  \includegraphics[width=5.0in]{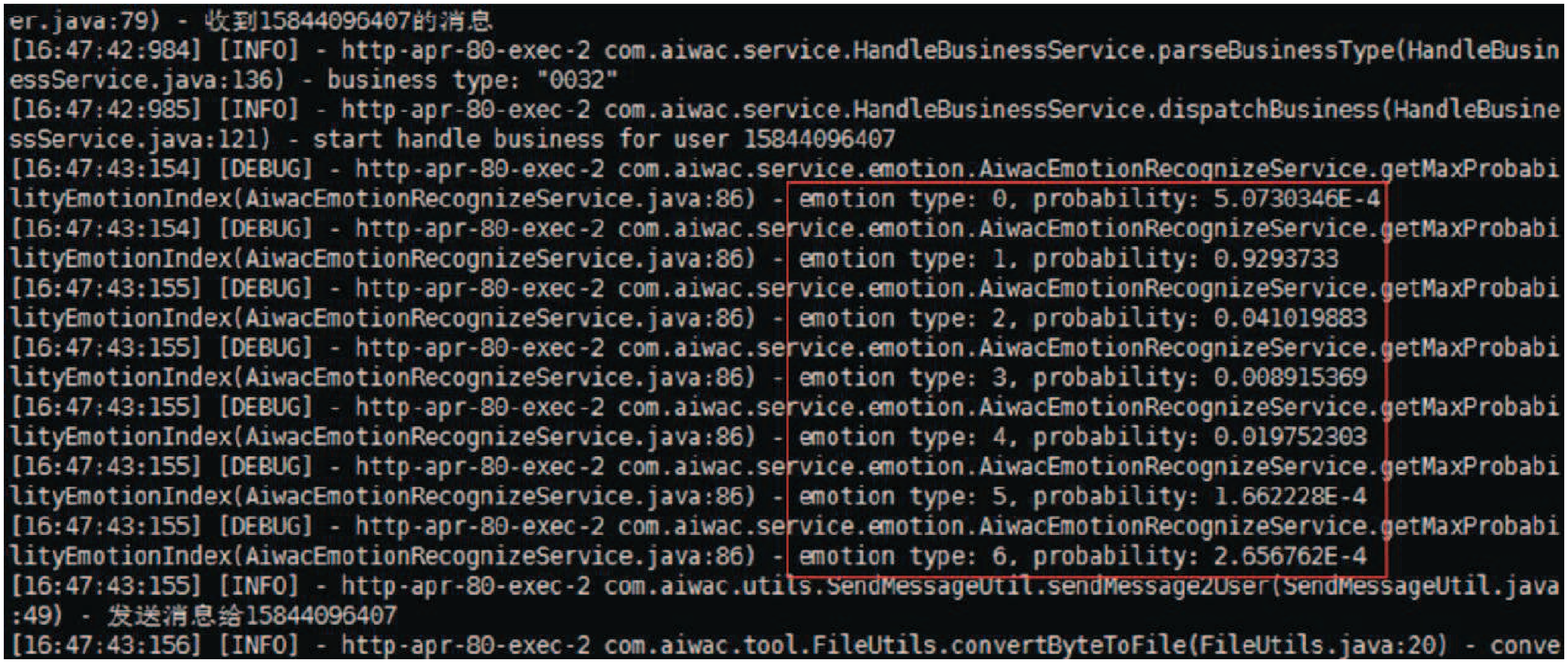}\\
  \caption{Results of Emotion Recognition}
  \label{fig04}
\end{figure*}

We have built a testbed platform of Fitbot to validate emotional communication mechanism, which includes smart phone terminal(edge server), Fitbot, and remote cloud with large data centers. The emotional data collected in the experiment are mainly voice emotion data, and Fitbot has a voice acquisition module. After the robot has collected a large number of users' emotional data, the robot transmits them to the edge server to be processed and annotated as unlabeled data, and then AI algorithm for emotion recognition is achieved in the remote cloud. According to the result of emotion recognition, as shown in Fig.~\ref{fig04}, robot and user interact with each other. Due to the dynamic change of the user's emotional state, the real-time emotional interaction is necessary, which requires that the communication link must have ultra-low latency. Another key element of emotional communication is that the AI emotion analysis algorithm deployed on the remote cloud(or in Fitbot) must have a high accuracy rate of user emotion recognition.

\section{WEARABLE AFFECTIVE ROBOT ALGORITHM} \label{sec.algorithmn}

\subsection{Affective Recognition Algorithm based on AIWAC Smart Box}

\begin{figure*}
\centering
\includegraphics[width=6in]{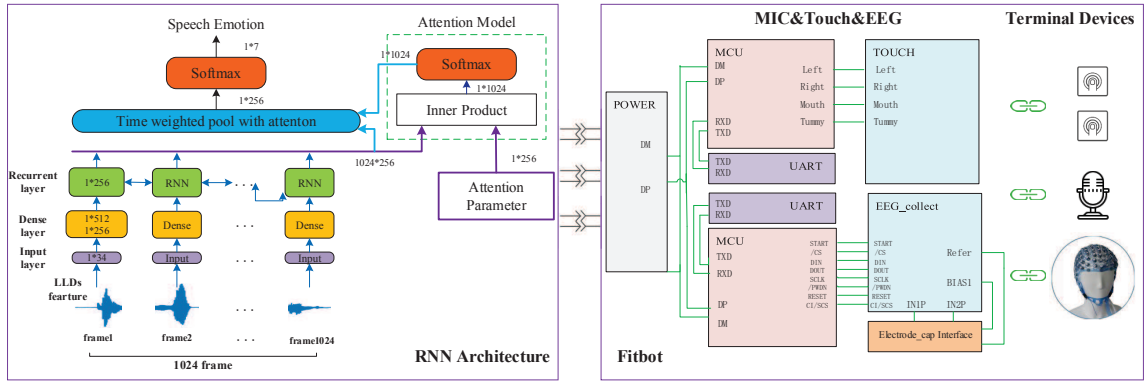}
\caption{The RNN algorithm for Fitbot Emotion Sensing}
\label{fig03}
\end{figure*}

The AIWAC smart box in a Fitbot has the ability of accurate voice affective cognition. Through daily voice interactions with a user, a Fitbot collects user's voice data and realizes voice affective analysis by using the attention based recurrent neural network algorithm. A recurrent neural network (RNN) can effectively remember the relevant feature information in a long context. By introducing an attention mechanism into the RNN algorithm framework, a new weight pooling strategy is introduced into the network, which puts special attention on the part expressing strong affective characteristics of the voice. The RNN architecture is shown in Fig.~\ref{fig03}.

\subsection{User Behavior Perception Based on Brain Wearable Device}

The brain wearable device can realize the perception of user's subtle expression change. On the one hand, the detection of these subtle behaviors enables the perception and analysis of user's behavior patterns; on the other hand, some high-efficiency human-computer interaction patterns may be perceived by virtue of these accurate behaviors.

The perception of user behaviors by a brain wearable device can be introduced by a blink detection example. We design an algorithm to realize such a detection.

\begin{figure}
\centering
\includegraphics[width=3.2in]{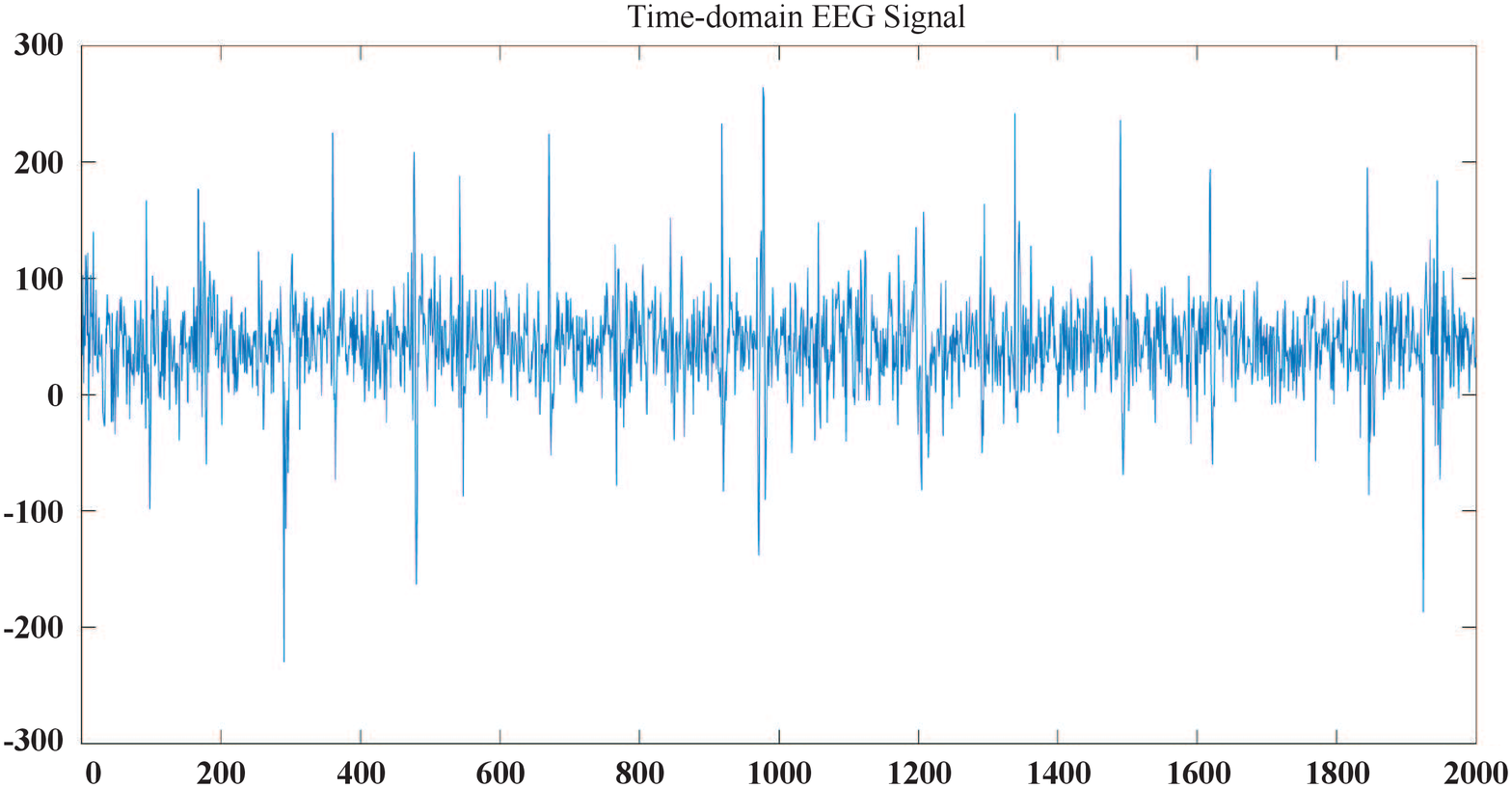}
\caption{Time-domain EEG Signal}
\label{fig0311}
\end{figure}

\begin{figure}
\centering
\includegraphics[width=3.6in]{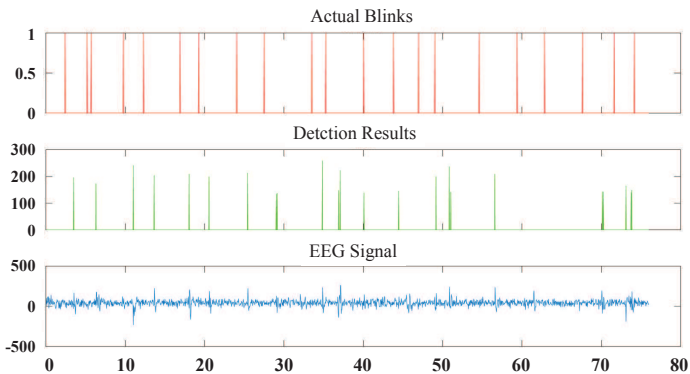}
\caption{Comparison between the actual blinks and detection results}
\label{fig0313}
\end{figure}

\textbf{Algorithm: Realization of blink detection based on the amplitude difference.}
The general process of Algorithm 1 is to determine the first order difference in the original EGG signal, then, the amplitude smoothing is performed, and lastly, the judgment on the smoothing result is conducted. The Algorithm 1 uses the time-domain data and draws the signal waveform diagram as shown in Fig.~\ref{fig0311}, wherein it can be seen that such a signal experiences a significant amplitude change during the blinking, which can be directly used to judge whether a user blinks or not. First, the first order difference in the original signal is determined to obtain the change rate information of the time-domain signal. To amplify the peak feature, the amplitude smoothing is used, i.e., the amplitude values lower than 150 are set to zero. By comparing the original signal with the result of the actual blink test and discrimination, it can be seen that in 20 blink tests, in 17 tests the blinks were detected accurately, as shown in Fig.~\ref{fig0313}.

\begin{table*}
\renewcommand{\arraystretch}{2}
\caption{Fitbot Application Service Scenarios}
\label{tab03} \centering
\begin{center}
\newcommand{\tabincell}[2]{\begin{tabular}{@{}#1@{}}#2\end{tabular}}

\begin{tabular}{|l|l|p{10cm}|}
\hline
\textbf{Scenes}	& \textbf{Items}	 & \textbf{Description} \\
\hline
\multirow{4}*{Home} & Smart Home &\multirow{4}{10cm}{Fitbot can accurately recognize the emotion of the people in an enclosed environment, so as to prevent the mental anomaly and safety problem under no supervision, which is like a butler tracking and accompanying a user 24/7.}\\
\cline{2-2}
~ & Stay-at-home children & ~\\
\cline{2-2}
~ & Empty-nest elderly & ~\\
\cline{2-2}
~ & Patient living alone & ~\\
\hline
\multirow{3}*{Hospital} & Outpatient clinic,  emergency treatment &\multirow{3}{10cm}{Fitbot reduces the occurrence of sudden diseases or medical troubles by recognizing and effectively intervening the emotions of patients and timely notifying the doctors and patient families. Fitbot is a doctor's intelligent assistant and the patient's guardian angel who not only can create a more humanized medical environment but also can reduce the tension between doctors and patients.}\\
\cline{2-2}
~ & Operating Room & ~\\
\cline{2-2}
~ & Inpatient Department & ~\\
\hline
\multirow{3}*{Education System} & Monitoring in the examination rooms &\multirow{3}{10cm}{\scriptsize Fitbot can accurately recognize the emotions of students, improving the education and teaching efficiency and reducing the occurrence of accidental injuries in school. Fitbot aims to realize three major purposes: expanding the education opportunities, improving the education quality, and reducing the education cost. It realizes the in-depth integration of information technology and education and teaching, forming a new talent-training mode which is multi-media, interactive, personalized, self-adaptive, and learner-centered.}\\
\cline{2-2}
~ & Campus security monitoring & ~\\
\cline{2-2}
~ & Remote teaching & ~\\
\hline
\multirow{4}*{Other public place} & Tourist Areas &\multirow{4}{10cm}{Aiming at other high-risk industries, crowd-intensive places, and areas with high incidence of crimes, in combination with the video big data collected by the security monitoring system, a Fitbot provides an intelligent emotion recognition algorithm and an intervention strategy to improve the security efficiency and reduce the occurrence of accidents. Good environment security management of the public places represents the civilization degree of a city or even a country.}\\
\cline{2-2}
~ & Chain stores & ~\\
\cline{2-2}
~ & Hotel & ~\\
\cline{2-2}
~ & Entertainment and business places & ~\\
\hline
\end{tabular}
\end{center}

\end{table*}

\section{LIFE MODELING OF WEARABLE AFFECTIVE ROBOT} \label{sec.life-modeling}

Through continuous in-depth and in-breadth data collection, a Fitbot builds the life modeling for users and conducts the long-term analysis and modeling of users' behaviors and emotions, which makes Fitbot have a deeper cognitive level on users and endows Fitbot with better user's experience of affective social interaction. The process of life modeling from three aspects is introduced below.

\subsection{Dataset Labeling and Processing}

In the process of data collection, the labeled data such as the user's physical examination reports and doctor's diagnostic records cannot be obtained in most cases. Therefore, a large amount of unlabeled data is collected in the unconscious interaction between the user and system. These data should be processed without disturbing a user.

After Fitbot collects the breadth and depth data of the user, it uses AI emotional communication to pass on data to the edge cloud for labeling and processing of data set~\cite{Touzri2016Efficient}~\cite{Duan2018}, and then it further transmits to the distal cloud and uses AI technology to have modeling analysis of the life modeling of the user, and feeds back to the user, so as to improve its cognition to the user's life modeling.

In the case of a small amount of data, the mathematical modeling is used to judge whether the unlabeled data should be added to the dataset. The labelless learning is used to decide whether the unlabeled data are added to the dataset or not based on the similarity measurement and taking into account the effect of data on the dataset after such data are added to the dataset. If such data influence positive to the entire dataset, the system will consider adding such data to the dataset. In addition, we also needs to consider the purity of the data. With the aim to improve the overall purity of the dataset, some unreliable data must be excluded because the ambiguous and low-value data will cause the spread of the error.


\subsection{Multi-dimensional Data Integration Modeling}

By using the hardware, the embedded control, the cloud platform of Big Data, and various deep learning algorithms, the various data are collected continuously through the IoT (Internet of Things) terminal devices, including the affective interaction robots, mobile phones (Android and iOS applications), etc. Through users' interaction with a social robot and his playing the affective cognitive games on a mobile phone, a lot of user data can be collected, including the pictures, environment background, voice, and text description of things to be done.

The text data are processed by the convolutional neural network, and the deep network is established mainly to extract the features from such unstructured data. In the case of the structured data, it is relatively easy to use the machine learning method directly, integrating multiple machine learning models including the decision-making tree, random forest, clustering algorithm, and others to obtain the robust models. These models can be used to predict users' emotions, preferences, and behaviors.

\subsection{Modeling of Associated Data Scenarios}

Since the dimension of user data is high and diversified, the inference of user data is not a simple one-to-one inference. There are many classes of image data, such as identity data. Therefore, after the person identity is determined, it is also needed to make a further inference from the data of the surrounding environment of a user. The data such as ``who'', ``where'', and ``when'', and even the sports data, denote the elements needed for inference. Since the information data are associated with the emotion in other aspects, the emotion cannot be inferred separately. Instead, the emotion should be inferred according to the information data in various aspects of a circumstance, and the reason for giving rise to such emotion should be inferred in a high degree of confidence. Therefore, in the modeling of associated data scenario, according to the behavioral fragments of a user, the events associated with emotion are extracted, and their associations with similar users and with other different users are explored.

In modeling, the key elements of multi-modal unstructured data are first extracted through the simplified mode. At this point, the emotions of a user are labeled in a multi-dimensional and multi-perspective way. Then, for the same user, many other users in different time periods and locations are labeled, and then the similarity of these users can be further explored.

The modeling of the associated data scenario not only can provide simple affective classification, but also can realize more complex scenario analysis to analyze the user's mental experiences, including the change and transformation of the user's emotional state in a certain situation, which makes a Fitbot more intelligent.

\section{APPLICATION}  \label{sec.application}

\subsection{Characteristics of Wearable Affective Robot}
\begin{itemize}
\item \textbf{Intelligence and Autonomy:} A real intelligent robot should have the intelligence and autonomy and should be able to refuse an unhealthy or an immoral request from a user. For instance, if a user requests from an affective robot to buy a junk food, the affective robot should refuse that request and explain the harmful effects resulting from the consumption of such a junk food, which will promote the establishment of trust between the affective robot and user, and improve the health surveillance effect on the user.
\item \textbf{Comfort and Portability:} The entity robot other than the virtual robot basically has limited movement ability, so it can accompany user with a space limitation. Its operation scenes are limited, and a user is often required to spend extra care to carry or maintain such an interactive robot. By virtue of the Wearable 2.0, a Fitbot is integrated into a wearable device in an natural way. The robot accompanies a user as a part of user's clothes, providing good portability to the robot. Meanwhile, a Fitbot provides convenient and various human-computer interaction modes, which make the user have good comfort.
\item \textbf{Diversified Human-Computer Interaction:} Current social robots are mainly based on voice interaction, while a Fitbot conveniently integrates tactile interaction on the basis of voice interaction, which expands the form of human-computer interaction, and provides available human-computer interaction means for some specific groups (such as the deaf-mute).
\item \textbf{Context Awareness and Personalized Tracking Service:} The collection of context data is conducive to the personalized services for a user. Generally, users having the social affective robots are also equipped with a smartphone. The context awareness data collected by a smartphone is closely related to a user and applies only to himself. A Fitbot can collect the context awareness data from the third-person (spectator) perspective, and can achieve more accurate identification of user activities. Such user data can be shared with multiple users such as family members or friends who are in the same room.
\end{itemize}

In addition, an affective robot can collect data from the first-person perspective and communicate with others on behalf of a user. Moreover, it can present the user's emotions in the form of a voice, a text, or an image, and convey others' information so as to make the interactions.

\subsection{Application Range}

On the basis of the affective recognition and interaction, a Fitbot imitates human behaviors and interacts with users through the voice interface. As the understanding on a user becomes deeper, a Fitbot will grow with the user and provide a user with a series of services. In Table \ref{tab03}, the application services provided by a Fitbot in four fields (home environment, health care, education, and public places) are listed.

\section{CONCLUSION}\label{sec.conclusion}
In this paper, a wearable affective robot equipped with cognitive computing named Fitbot is proposed. Joining the AIWAC smart box, smart tactile device, and brain wearable device, the complete system of a Fitbot is built. A Fitbot can realize the multi-modal data perception. Based on the perceived multi-modal data of a user, a Fitbot can cognize the user's emotions. Human emotions denote the external manifestation of a user's intentions, which enables a Fitbot to understand the behavioral motivations that are behind the user's emotions. Finally, the application service scenarios and some open questions of a Fitbot are discussed.

\section*{Acknowledgement}
This work was supported by the National Natural Science Foundation of China (under Grant No. 61875064, U1705261, 61572220), the National Key R\&D Program of China (2018YFC1314600), Director Fund of WNLO, the Fundamental Research Funds for the Central Universities (HUST: 2018KFYXKJC045), the National 1000 Talents Program, China, the Hubei Provincial Key Project under grant 2017CFA051, and the Applied Basic Research Program through Wuhan Science and Technology Bureau under Grant 2017010201010118.

\bibliographystyle{IEEEtran}

\begin{IEEEbiography}
[{\includegraphics[width=1in,height=1.25in,clip,keepaspectratio]{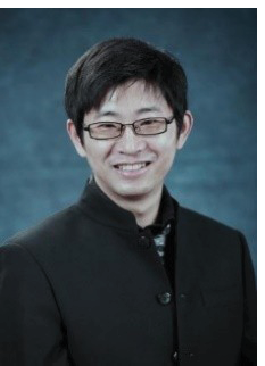}}]{Min Chen} is a full professor in School of Computer Science and Technology at Huazhong University of Science and Technology (HUST) since Feb. 2012. He is the director of Embedded and Pervasive Computing (EPIC) Lab at HUST. He is Chair of IEEE Computer Society (CS) Special Technical Communities (STC) on Big Data. He was an assistant professor in School of Computer Science and Engineering at Seoul National University (SNU). He worked as a Post-Doctoral Fellow in Department of Electrical and Computer Engineering at University of British Columbia (UBC) for three years. Before joining UBC, he was a Post-Doctoral Fellow at SNU for one and half years. He received Best Paper Award from QShine 2008, IEEE ICC 2012, ICST IndustrialIoT 2016, and IEEE IWCMC 2016. He serves as technical editor or associate editor for IEEE Network, Information Sciences, Information Fusion, and IEEE Access, etc. He served as a leading Guest Editor for IEEE Wireless Communications, IEEE Network, and IEEE Trans. Service Computing, etc. He is a Series Editor for IEEE Journal on Selected Areas in Communications. He is Co-Chair of IEEE ICC 2012-Communications Theory Symposium, and Co-Chair of IEEE ICC 2013-Wireless Networks Symposium. He is General Co-Chair for IEEE CIT-2012, Tridentcom 2014, Mobimedia 2015, and Tridentcom 2017. He is Keynote Speaker for CyberC 2012, Mobiquitous 2012, Cloudcomp 2015, IndustrialIoT 2016, Tridentcom 2017 and The 7th Brainstorming Workshop on 5G Wireless. He has more than 300 paper publications, including 200+ SCI papers, 80+ IEEE Trans./Journal papers, 25 ESI highly cited papers and 9 ESI hot papers. He has published eight books: OPNET IoT Simulation (2015), Big Data Inspiration (2015), 5G Software Defined Networks (2016) and Introduction to Cognitive Computing (2017) with HUST Press, Big Data: Related Technologies, Challenges and Future Prospects (2014) and Cloud Based 5G Wireless Networks (2016) with Springer, Cognitive Computing and Deep Learning (2018) with China Machine Press, and Big Data Analytics for Cloud/IoT and Cognitive Computing (2017) with Wiley. His Google Scholars Citations reached 13,400+ with an h-index of 58. His top paper was cited 1500+ times. He is an IEEE Senior Member since 2009. He was selected as Highly Cited Research at 2018. He got IEEE Communications Society Fred W. Ellersick Prize in 2017. His research focuses on cognitive computing, 5G Networks, embedded computing, wearable computing, big data analytics, robotics, machine learning, deep learning, emotion detection, IoT sensing, and mobile edge computing, etc.
\end{IEEEbiography}

\begin{IEEEbiography}
[{\includegraphics[width=1in,height=1.25in,clip,keepaspectratio]{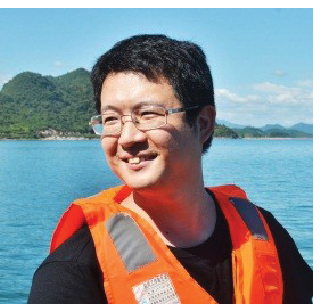}}]{Jun Zhou} is a full professor in Wuhan National Laboratory for Optoelectronics (WNLO) at Huazhong University of Science and Technology (HUST) since 2009. He is the deputy director of WNLO. He received his Bachelor Degree (2001) in materials physics and Ph.D. degree (2007) in materials physics and chemistry from the Sun Yat-sen University. During 2005-2006, He was a visiting student at school of materials science and engineering, Georgia institute of technology. During 2007-2009, he served as a research scientist in the Wallace H. Coulter department of biomedical engineering and school of materials science and engineering, Georgia institute of technology. He has published over 130 peer reviewed journal papers, including 3 ESI hot papers and 29 ESI highly cited papers. All of the papers have been cited over 11000 times. His H-index is 52 and he is one of the top 0.1\% highly cited author in the Royal Society of Chemistry Journals in 2014. He has organized 6 conferences, and delivered over 30 invited talks in conferences. He has awarded the National Natural Science Award of Chinese government (second prize) for the Year of 2016, Natural Science Award of Ministry of Education Department of China (first prize) for the Year of 2015, and the Excellent Doctoral Dissertation of China for the Year of 2009. He also has been awarded Excellent Youth fund of National Natural Science Foundation of China on year of 2013, enrolled in National Program for Support of Top-notch Young Professionals for the Year of 2014, Youth project for ``Cheung Kong Scholars programme'' of Ministry of Education Department of China for the Year of 2015. His research focuses on energy harvesting from environmental and flexible electronics.
\end{IEEEbiography}

\begin{IEEEbiography}
[{\includegraphics[width=1in,height=1.25in,clip,keepaspectratio]{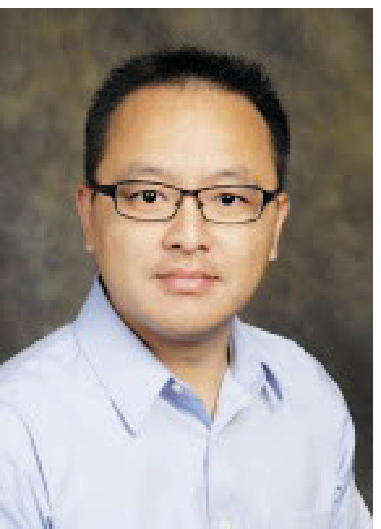}}]{Guangming Tao} is a Professor at Wuhan National Laboratory for Optoelectronics and the School of Optical and Electronic Information at Huazhong University of Science and Technology (HUST). He is the director of Center of Advanced Functional Fibers (CAFF) and the director of Man-Machine Lab (2M lab) at HUST. He received his Ph.D. degree (2014) in optics from the University of Central Florida. He was a Research Scientist/Sr. Research Scientist at The College of Optics \& Photonics (CREOL), University of Central Florida from 2014 to 2017. He was a visiting scholar at Chinese Academy of Science (2007-2008), the Massachusetts Institute of Technology (2012), and Centre national de la recherche scientifique (2017). Dr. Tao has published about 35 scientific papers, holds 7 U.S. and foreign patents, has given in excess of 45 invited lectures/colloquia or keynote talk, and has co-organized more than 10 national and international conferences and symposia, including Symposium SM2 (Advanced multifunctional fibers and textiles) at 2017 Spring MRS Meeting, Symposium J (Multifunctional and multimaterial fibers) at 2017 International Conference on Advanced Fibers and Polymer Materials, etc. He has years of research experience in optical sciences and engineering in academia, industry, and government institutes with expertise in the areas of functional fibers, smart fabric, man-machine interactions, specialty optical fibers and in-fiber nano-fabrication.
\end{IEEEbiography}

\begin{IEEEbiography}
[{\includegraphics[width=1in,height=1.25in,clip,keepaspectratio]{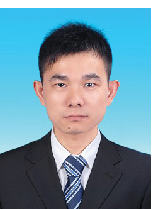}}]{Jun Yang} received Bachelor and Master degree in Software Engineering from Huazhong University of Science and Technology (HUST), China in 2008 and 2011, respectively. Then, he got his Ph.D degree at School of Computer Science and Technology, HUST, on June 2018. Currently, he works as a postdoctoral fellow at Embedded and Pervasive Computing (EPIC) Lab in School of Computer Science and Technology, HUST. His research interests include cognitive computing, software intelligence, Internet of Things, cloud computing and big data analytics, etc.
\end{IEEEbiography}

\begin{IEEEbiography}
[{\includegraphics[width=1in,height=1.25in,clip,keepaspectratio]{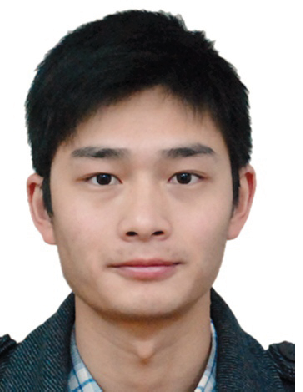}}]{Long Hu} is a lecturer in School of Computer Science and Technology at Huazhong University of Science and Technology (HUST). He has also received his Doctor, Master and B.S. degree in HUST. He is the Publication Chair for 4th International Conference on Cloud Computing (CloudComp 2013). Currently, his research includes 5G Mobile Communication System, Big Data Mining, Marine-Ship Communication, Internet of Things, and Multimedia Transmission over Wireless Network, etc.
\end{IEEEbiography}


\end{document}